\documentclass[ prl, amsmath,twocolumn,superscriptaddress]{revtex4-1}

\usepackage{verbatim}
\usepackage{graphicx}

\begin{document}

\title{Multimode Optomechanics as a Prototype of Mediated Interactions}
\author{L.~F.~Buchmann}
\affiliation{Department of Physics, University of California, Berkeley, California 94720, USA}
\author{D.~M.~Stamper-Kurn}
\affiliation{Department of Physics, University of California, Berkeley, California 94720, USA}
\affiliation{Materials Sciences Division, Lawrence Berkeley National Laboratory, Berkeley, CA 94720, USA}
\pacs{blabla}
\begin{abstract}
We theoretically investigate two quantum modes interacting via local couplings to a dissipative field. Our model considers two mechanical modes with distinct frequencies coupled optomechanically to the same cavity mode. The dissipative cavity field mediates the interaction between the mechanical modes but also leads to decoherence of the mechanical oscillators. Depending on the ratio between effective interaction strength and dissipation rate, which can be chosen via the pump detuning, the interaction assumes a quantum mechanical or classical character. For any cavity decay, there is a regime where the two mechanical modes interact in a non-classical way, which leads us to conclude that optomechanical systems can serve as a model to experimentally study the effects of long-range interactions mediated by classical or quantum-mechanical fields. 
\end{abstract}
\maketitle
The emergence of a classical world from quantum theories remains a challenge to our understanding of physics. One important mechanism through which microscopic interactions manifest themselves in macroscopic effects is long-range forces, such as the Coulomb interaction or gravitation. In microscopic descriptions, Lorentz invariance requires these forces to be the result of modes coupling locally to ``force carriers'', such as the photon or the hypothetical graviton. Tracing out the mediating field -- whose excitations are referred to as ``virtual'' since they cannot be directly measured --  yields the effective long-range interaction. \\
According to a recently proposed model, two bodies experiencing a long-range interaction mediated by a classical, rather than a quantum field, will experience additional random forces due to the carrier's inability to create non-classical correlations~\cite{kafri1}. This additional noise causes the position variances of the two interacting modes to increase at a rate larger than twice the effective coupling strength. In Ref.~\cite{kafri2}, the same bound was found starting from a different model, where the description of a classical long-range interaction was based on quantum measurement theory. A field mediating a long-range interaction is a channel carrying information about the participating bodies. The authors of \cite{kafri1,kafri2} draw the line between classcially and quantum-mechanically interacting bodies at the ability of the force-carrier to transfer quantum information. If, for example, one of the bodies is in a in a non-classical superposition state, a quantum mechanical force-carrier would faithfully convey this and -- in the appropriate basis -- also be in a superposition state. A classical channel, however, could not transfer this quantum information and thus must carry additional noise affecting both participating bodies. 
Besides the fundamental question of a possible quantum character of gravitation, the classicality of an interaction is of practical importance to technological applications of quantum mechanics, most prominently quantum computation.
\\
In this Letter, we show that cavity optomechanics~\cite{review} makes the study of such fundamental questions feasible by providing an experimentally realizable and highly controllable model system. The control over quantum aspects of macroscopic oscillators using optomechanical methods has been pointed out earlier~\cite{nonclassicalOM, MAQRO, tunneling,nonclassicalOM2, Yanbei} and we already witnessed the unmatched sensitivity of optomechanical force detection~\cite{forcesensitivity} and the observation of non-classical behavior\cite{OMsqueezing, Cleland, ultracold, sidebandasym, forcesensitivity}. We propose to use optomechanical systems as simulators of mediated interactions and show how the interaction between distinct mechanical modes mediated by the field of a cavity can be tuned to cover classical and non-classical regions; the detuning of an appropriate pump from the cavity resonance being a convenient control parameter. The interaction can take on a non-classical character irrespective of the cavity's quality factor, although some quantum-mechanical effects vanish if the electromagnetic decay rate becomes comparable to the average frequency of the mechanical modes.\\
Our theoretical treatment does not implicitly rely on an optomechanical system and remains valid for any interaction mediated by a dissipative field that can be modeled as a harmonic oscillator. We chose to illustrate the results in the context of optomechanics because of the tremendous technical progress in this field~\cite{review}, which allows to study mediated interactions in a broad range of regimes. Moreover, our results are of practical interest in multimode optomechanics~\cite{multimodematt, multimodeOM1, multimodeOM2, multimodeOM3, multimodeOM4} since they provide a prescription to resonantly couple mechanical modes with frequencies orders of magnitude apart, while coupling mechanical modes has so far only be demonstrated between near-degenerate modes~\cite{multimodeOM5,multimodeOM6,multimodeOM7}. Such a coupling can be used to optomechanically cool low-frequency mechanical modes, extend the sensitivity of mechanical sensors or use several mechanical elements in quantum computing~\cite{computing}.

\begin{figure}
\includegraphics[width=0.45\textwidth]{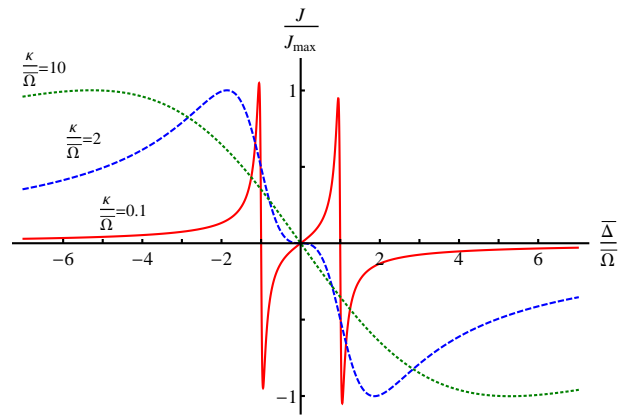}
\caption{Normalized coupling strength as a function of central detuning for different side-band resolutions.}
\label{couplingstrengthfig}
\end{figure}

\begin{figure*}[h!t]
\includegraphics[width=0.35\textwidth]{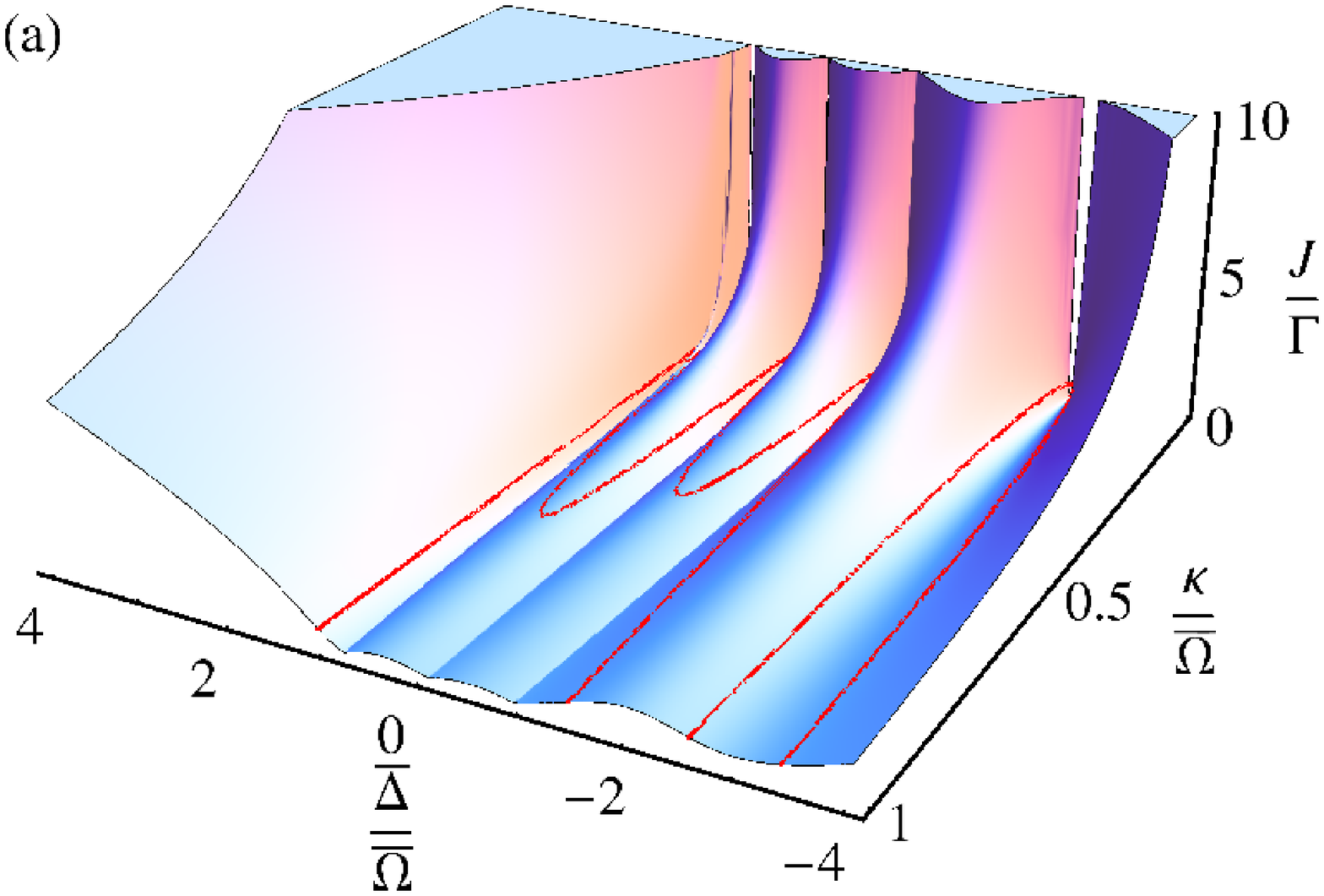}
\hspace{1.5cm}
\includegraphics[width=0.35\textwidth]{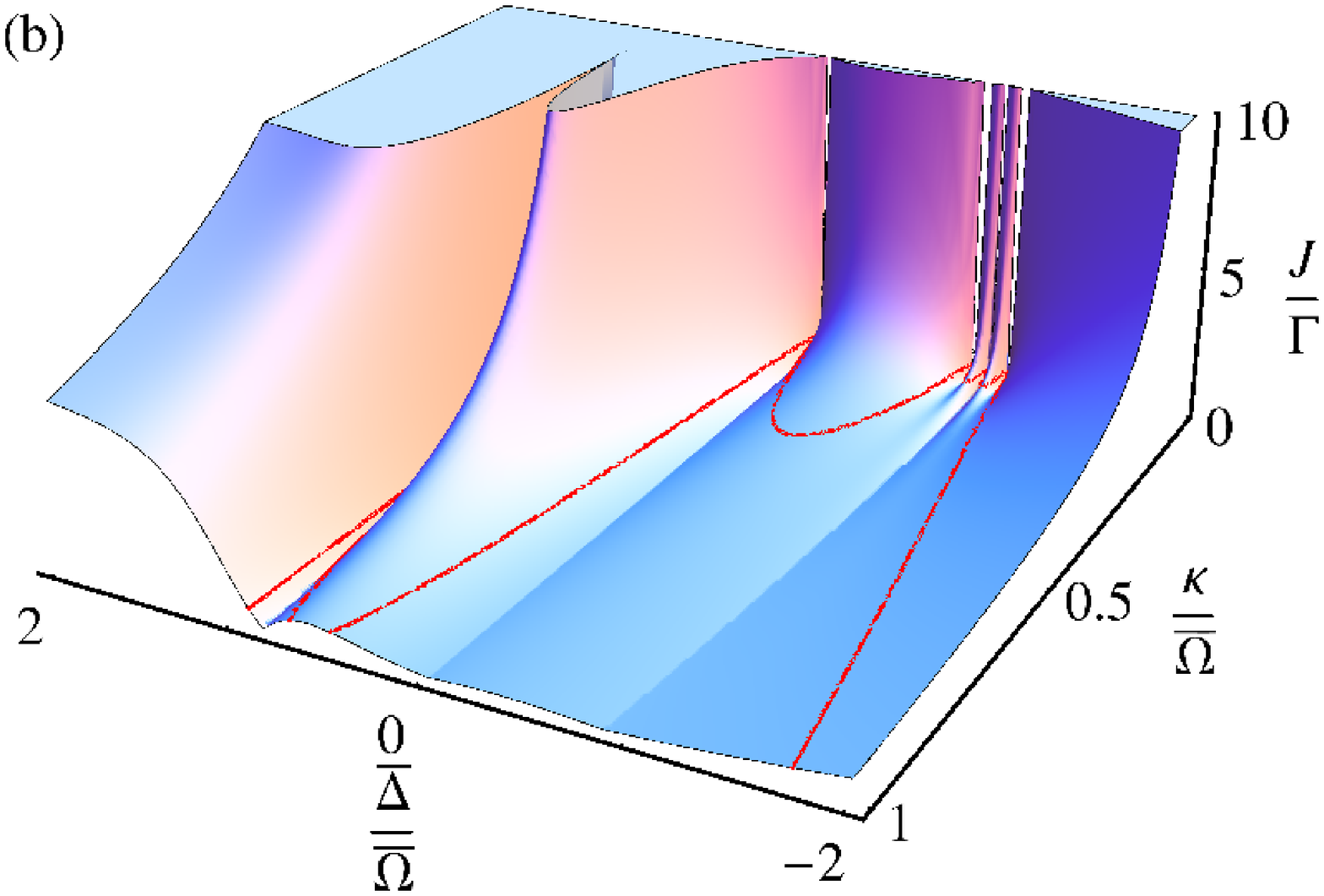}
\vspace{0.5cm}
\includegraphics[width=0.35\textwidth]{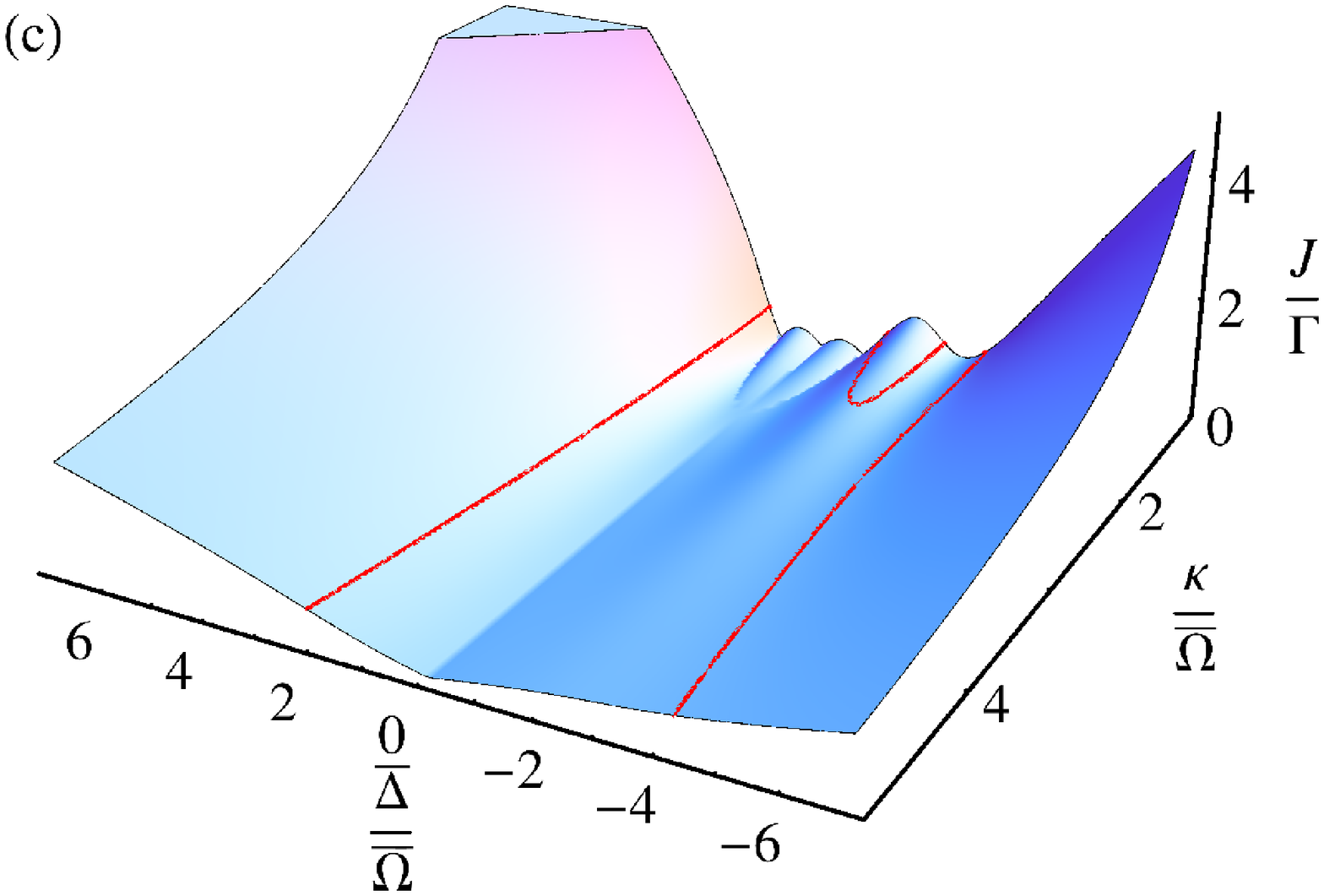}
\hspace{1.5cm}
\includegraphics[width=0.35\textwidth]{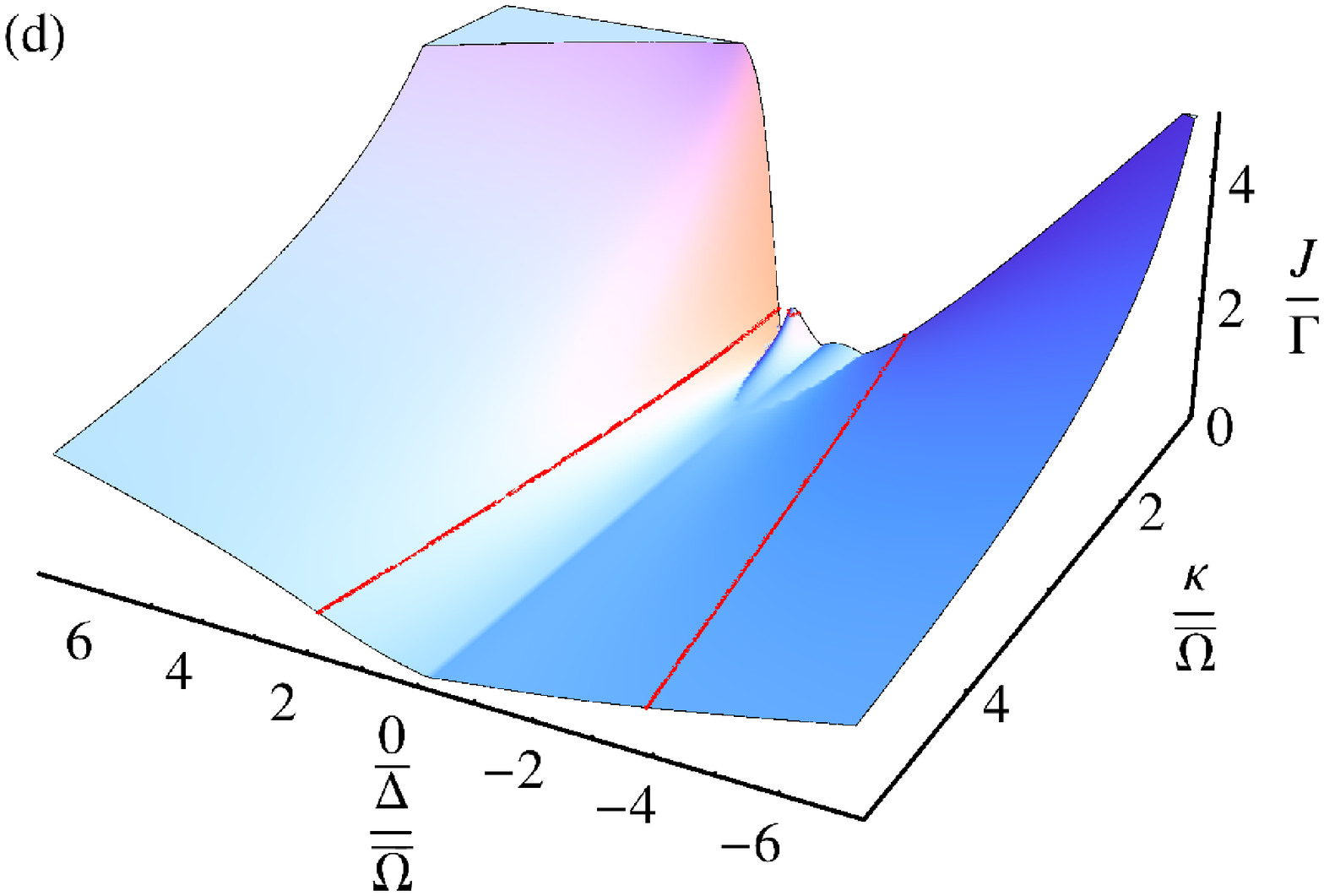}
\caption{Ratio of effective interaction strength to total excess noise due to the dissipative nature of the mediating field for different side-band resolution regimes (top and bottom rows) and frequency differences between the mechanical modes (left and right columns). For panels (a) and (c) we have $\delta\omega/\bar{\Omega}=1.9$, while for (b) and (d) $\delta\omega/\bar{\Omega}=0.1$. The line separates classical (below) and quantum (above) regimes.}
\label{couplingratio}
\end{figure*}

Consider an optomechanical system involving two mechanical modes with distinct frequencies, $\omega_1\neq\omega_2$, and annihilation operators $\hat{b}_j, j=1,2$. Both modes interact with a single electromagnetic cavity mode $\hat{a}$ with resonant frequency $\omega_c$. In solid-state realizations of optomechanics different modes are naturally present~\cite{multimodeOM5,multimodeOM6,multimodeOM7} and have been shown to exhibit similar quality factors as the commonly used fundamental mode \cite{regalmultimode, mukundmultimode}. In realizations involving ultracold atoms, distinct mechanical modes can be prepared by trapping atomic clouds in a superlattice \cite{coldatomsmultimode}. 
In the lab frame, the Hamiltonian of the system is given by 
\begin{equation}
\mathcal{H}=\mathcal{H}_\mathrm{opt}+\mathcal{H}_\mathrm{m}+\mathcal{H}_\mathrm{I}+\mathcal{H}_\kappa
\end{equation}
with 
\begin{align}
\mathcal{H}_\mathrm{opt}&=\omega_c\hat{a}^\dag\hat{a}+\eta(t)\hat{a}^\dag+\eta^*(t)\hat{a}
\end{align}
\begin{align}
\mathcal{H}_\mathrm{m}&=\sum_{j=1}^2\omega_j\hat{b}_j^\dag\hat{b}_j
\end{align}
\begin{align}
\mathcal{H}_\mathrm{I}&=\sum_{j=1}^2g_j\hat{a}^\dag\hat{a}(\hat{b}_j^\dag+\hat{b}_j),
\end{align}
and $\mathcal{H}_\kappa$ accounts for dissipation of the cavity field at a rate $\kappa$ to a reservoir at zero temperature. We have set $\hbar=1$, droppped constant terms and defined the pumping rate $\eta$ and the single-photon optomechanical couplings $g_j$. The coupling of each oscillator to its thermal reservoir is left out because we want to focus on decoherence due to the dissipative nature of the field mediating the interaction and discuss the individual heat-baths later.
The density operator $\rho$ satisfies the master equation
\begin{equation}\label{mastereq1}
\frac{d\rho}{dt}=-i[\mathcal{H},\rho]+\kappa\mathcal{L}(\hat{a})\rho,
\end{equation}
with the Lindblad superoperator
\begin{equation}
\mathcal{L}(\hat{a})\rho=\hat{a}\rho\hat{a}^\dag-\frac{1}{2}\hat{a}^\dag\hat{a}\rho-\frac{1}{2}\rho\hat{a}^\dag\hat{a}.
\end{equation}
Tracing out the electromagnetic system will yield an effective master equation coupling the mechanical modes. For the coupling to become resonant we need the cavity field to oscillate at the frequency difference of the mechanical oscillators, which can be achieved by driving the cavity with a bi-chromatic pump,
\begin{equation}
\eta(t)=\eta_1e^{-i\omega_{L1}t}+\eta_2e^{-i\omega_{L2}t},
\end{equation}
and the pump frequencies satisfying 
\begin{equation}\label{frequencycondition}
\omega_{L1}-\omega_{L2}=\omega_2-\omega_1.
\end{equation}
Similarly to non-linear optics, the beat-note between the two drives will make up for the energy difference of phonons in each oscillator. If one were interested in creating entanglement or phase conjugation~\cite{phasecon}, a different choice for the pump frequencies would be more prudent, but for our purposes of mediated, linear interaction, condition (\ref{frequencycondition}) is appropriate. 
To eliminate the light field, we first unitarily displace the cavity field operator by a time-dependent complex function: $\hat{a}\to\alpha(t)+\hat{a}$. If we choose 
\begin{equation}
\alpha(t)=\alpha_1e^{-i\omega_{L1}t}+\alpha_2e^{-i\omega_{L2}t},
\end{equation}
with 
\begin{equation}
\alpha_j(t)=\frac{-i\eta_j}{\kappa/2+i\Delta_j},
\end{equation}
where $\Delta_j=\omega_c-\omega_{Lj}$, all source terms for $\hat{a}$ cancel and it only carries the quantum fluctuations of the cavity field. Without loss of generality, we may choose the phases of $\eta_j$ such that both $\alpha_j$ are real. For the small couplings characteristic of optomechanics~\cite{review}, we can neglect the interactions quadratic in $\hat{a}$ and after absorbing the free evolutions in the operators the Hamiltonian in Eq.~(\ref{mastereq1}) becomes
\begin{equation}
\sum_{j=1}^2g_j\left(\alpha(t)\hat{a}^\dag e^{i\omega_c t}+\alpha^*(t)\hat{a}e^{-i\omega_ct}\right)\left(\hat{b}_j^\dag e^{i\omega_jt}+\hat{b}_je^{-i\omega_jt}\right).
\end{equation}
To perform the partial trace over electromagnetic degrees of freedom we restrict the optical Hilbert space to the optical vacuum and first-order coherences. By virtue of the displacement operation this is a reasonable approximation and corresponds to a first-order perturbation theory in the optomechanical coupling strength. The tracing out of the cavity field is straight-forward but yields lengthy intermediate expressions, which are given in the supplementary material~\cite{supplement}. The resulting effective master equation reads
\begin{align}
\frac{d\rho_\mathrm{m}}{dt}=&-i[\mathcal{H}_\mathrm{eff},\rho_\mathrm{m}]
+\bar{\Gamma}\left[(\bar{n}+1)\mathcal{L}(\hat{B})\rho_\mathrm{m}+\bar{n}\mathcal{L}(\hat{B}^\dag)\rho_\mathrm{m}\right]
\nonumber\\
&+\sum_{j=1}^2\Gamma_j\left[(\bar{n}_j+1)\mathcal{L}(\hat{b}_j)\rho_\mathrm{m}+\bar{n}_j\mathcal{L}(\hat{b}_j^\dag)\rho_\mathrm{m}\right],\label{effmastereq}
\end{align}
where we introduced the collective mode: 
\begin{equation}\label{colmode}
\hat{B}=\sqrt{\frac{g_1}{g_2}}\hat{b}_1+\sqrt{\frac{g_2}{g_1}}\hat{b}_2,
\end{equation}
and the effective Hamiltonian is
\begin{equation}\label{effhamiltonian}
\mathcal{H}_\mathrm{eff}=\sum_{j=1}^2\delta\Omega_j\hat{b}_j^\dag\hat{b}_j+J\hat{b}_1^\dag\hat{b}_2+J^*\hat{b}_1\hat{b}_2^\dag. 
\end{equation}
The frequency shift $\delta\Omega_j$ is the sum of the optical springs arising from the interaction of oscillator $j$ with the two cavity pumps. Its explicit form can be found in the literature~\cite{review} and its effect can be incorporated in the operators $\hat{b}_j$.
The other parameters are found to be
\begin{subequations}\label{parameters}
\begin{align}
J=&G_1G_2\mathrm{Im}\left[\frac{\kappa+2i\bar{\Delta}}{(\kappa/2+i\bar{\Delta})^2+\bar{\Omega}^2}\right],\label{effcoupling}\\
\Gamma_j=&\frac{2G_j^2\kappa\bar{\Delta}(\bar{\Omega}\pm\delta\omega)}{(\kappa^2/4+(\bar{\Omega}\pm\delta\omega)^2-\bar{\Delta}^2)^2+\kappa^2\bar{\Delta}^2},\label{gammaj}\\
\bar{\Gamma}=&\frac{2G_1G_2\kappa\bar{\Delta}\bar{\Omega}}{(\kappa^2/4+\bar{\Omega}^2-\bar\Delta^2)^2+\kappa^2\bar{\Delta}^2},\label{gammabar}\\
\bar{n}_j=&\frac{\kappa^2}{16\bar{\Delta}(\bar{\Omega}\pm\delta\omega)}+\frac{\bar{\Delta}}{4(\bar\Omega\pm\delta\omega)}+\frac{\bar\Omega\pm\delta\omega}{4\bar\Delta}-\frac{1}{2},\label{nj}\\
\bar{n}=&\frac{\kappa^2}{16\bar\Delta\bar\Omega}+\frac{\bar\Delta}{4\bar\Omega}+\frac{\bar\Omega}{4\bar\Delta}-\frac{1}{2},\label{nbar}
\end{align}
\end{subequations}
where the lower (upper) sign in $\Gamma_j$ and $\bar{n}_j$ has to be taken for $j=1$  $(j=2)$ and we we introduced the dressed couplings $G_j=g_j\alpha_j$ with $\alpha_1=\alpha_2$ as well as the average frequency $\bar\Omega=\frac{\omega_1+\omega_2}{2}$ , the frequency difference $\delta\omega=\omega_1-\omega_2$ and the central detuning $\bar\Delta=\frac{\Delta_1+\Delta_2}{2}$ to simplify notation. \\
Eq.~(\ref{effmastereq}) describes the evolution of two oscillators coherently exchanging excitations at a rate $|J|$, coupled to three heat-baths; one for each mode separately and one coupling to the collective mode (\ref{colmode}). The effective interaction (\ref{effhamiltonian}) can extend the bandwidth of mechanical oscillators used as sensors or allow the cooling of a low-lying mechanical mode using a high-frequency mechanical oscillator, but we will not discuss these effects further.
The single-mode baths characterized by Eqs. (\ref{gammaj}) and (\ref{nj}) are known from single-mode optomechanics \cite{review, ash}. They describe optomechanical damping and amplification corresponding to negative and positive effective temperatures for blue and red detuned pumps respectively. The collective bath with parameters given in Eqs. (\ref{gammabar}), (\ref{nbar}) is a consequence of the electromagnetically mediated interaction and its coupling and temperature depend only on the collective parameters $\bar{\Delta}$ and $\bar{\Omega}$. In the limit $\kappa\to 0$ the dissipation vanishes and the interaction between the mechanical modes becomes unitary. For non-vanishing values of $\kappa$, the three baths need to be taken into account to describe the system. It is straightforward to add thermal baths of the two oscillators to Eq.~(\ref{effmastereq}) and the consequences of doing so are intuitive. In principle the effects of coupling to external heat-baths can always be made negligible by increasing the intracavity photon number, since the couplings due to cavity dissipation all scale with this quantity, while the thermal coupling will remain the same. Of course there are practical limits to such a solution, and it is desirable to conduct experiments in cold environments.\\
In any interaction of two bodies mediated by a quantum or classical field, the interaction in (\ref{effhamiltonian}) will be the first order contribution, corresponding to the two oscillators exchanging one excitation~\cite{Note1}. 
The dissipation terms are an unavoidable byproduct of the dissipative nature of the field mediating the interaction between the mechanical oscillators. Increasing the intracavity photon number will result in a stronger coherent coupling rate $|J|$ but also in higher dissipation rates, as each intracavity photon contributes to the mediation of the force between the oscillators, but also loses a certain fraction due to cavity decay. The loss of unitarity comes with information about the oscillators leaving the cavity and being accessible to measurement. This renders the exchanged excitations only partially ``virtual'' and the ratio between the unitary and dissipative processes is what determines the classicality of the mediated interaction. \\
Before we investigate this ratio further, we point out a quantum-mechanical interference effect present in this system. The behavior of the magnitude of the coherent coupling as a function of central detuning is plotted in Fig.~\ref{couplingstrengthfig} for different values of the side-band resolution $\kappa/\bar{\Omega}$. 
For decay rates $\kappa$ smaller than $2\bar\Omega$, there are two non-zero detunings for which the coherent transfer of excitations vanishes. This is a result of destructive interference between two distinct pathways for the exchange process. From Eq.~(\ref{frequencycondition}) it follows trivially that $\omega_{L1}+\omega_1=\omega_{L2}+\omega_2$ and $\omega_{L1}-\omega_2=\omega_{L2}-\omega_1$, i.e. two side-bands of each pump overlap. A cavity photon can deposit an excitation from oscillator 1 in oscillator 2 by first absorbing a phonon from oscillator 1 and then emitting it to oscillator 2, or vice versa by initial emission to oscillator 2 and then absorption from oscillator 1. Each of these processes has a different complex amplitude and the two amplitudes can coherently add up to zero. When the decay rate of the cavity becomes comparable to the mechanical frequency, the coherence time of the cavity field is not sufficient to interfere the two processes and the effect disappears.\\
One would be inclined to think it is this transition which turns the mediated interaction from a quantum-mechanical exchange of excitations to a classical spring between two oscillators. However, following \cite{kafri1, kafri2} we draw the line between classical and non-classical interacting modes where the mediated interaction loses the ability to entangle the two oscillators. This transition occurs when the decoherence rate 
\begin{equation}
\Gamma=\Gamma_1\bar{n}_1+\Gamma_2\bar{n}_2+2\bar{\Gamma}\bar{n}
\end{equation}
becomes larger than twice the effective coupling strength $|J|$, i.e. 
\begin{equation}
\xi\equiv\frac{|J|}{\Gamma}\le\frac{1}{2}.
\end{equation}
From Eqs. (\ref{parameters}) we calculate $\xi$ and plot it as a function of central detuning and side-band resolution for mechanical oscillators with similar and widely different frequencies in Fig.~\ref{couplingratio} (a),(c) and (b),(d) respectively. 
Classical and non-classical regimes of interaction are separated by a line, with the regions below the line corresponding to classical interaction. In the resolved side-band regime (top panels) $\xi$ is typically larger for central detunings around the cavity resonance. However, even for large cavity decay $\kappa$ the mediated interaction becomes quantum mechanical for large detunings $\bar\Delta$. From Eqs. (\ref{effcoupling}) -- (\ref{gammabar}) we find that for $\bar\Omega\ll\bar\Delta,\kappa$, the dissipative  couplings $\Gamma_j$ and $\bar\Gamma$ decrease faster than the effective coupling rate as a function of $\bar\Delta$, the latter following a dispersive curve. Far from $\bar{\Delta}=0$, $\xi$ is approximately quadratic in $\bar{\Delta}$ and can assume arbitrarily large values. In particular, the coupling can in principle overcome limitations by the thermal baths of each oscillator and there are no fundamental limits preventing a non-classical interaction of oscillators at any temperature, although practical limitations naturally apply. 
\\
From a practical point of view a high value of $\xi$ is desirable if one were interested in using mechanical oscillators as resources for quantum computation~\cite{computing}, since a value below $1/2$ indicates the inability to create entanglement between two mechanical modes or transfer nonclassical states. The actual coherent exchange rates depend naturally on the particular realization; 
Eq. (\ref{effcoupling}) shows that for ideal detuning the effective coupling is of the same order of magnitude as the optical spring and damping/amplification rates. This puts the effective coupling $J$ in the MHz range for optomechanical crystal cavities~\cite{sidebandasym}, and in the kHz regime for membrane-in-the-middle setups~\cite{multimodeOM5} or optomechanical systems involving ultracold atoms~\cite{review}. The influx of noise depends highly on the sideband resolution as well as the central detuning. For systems in the resolved sideband limit, the coupling can be orders of magnitude stronger than the additional noise for detunings close to cavity resonance. For systems with large optical decay $\kappa$, however, it is preferrable to be far away from the mechanical sidebands and compensate weak couplings with a high intracavity photon number.

In summary, we have provided a protocol to resonantly couple mechanical modes with arbitrary frequency differences in cavity optomechanics. In addition to the practical applications, we find this system to be a versatile prototype for the study of long-range interactions mediated by the shared cavity field. Since the decoherence due to the dissipation of a mediating field drops off faster than the \mbox{dispersion-like} effective coupling, the mediated interaction can take on non-classical characteristics even outside the resolved side-band regime for large enough detunings of the pump from the cavity resonance.

This work was supported by the SNSF, AFOSR and NSF. LFB would like to thank D.~Kafri and Y.~Chen for useful discussions.

\end{document}


\section*{Supplementray Material: Derivation of the Effective Master Equation}
Denoting Fock states of the optical Hilbert space with $|j\rangle_\mathrm{o}$ and introducing the mechanical quadrature operator 
\begin{equation}
\hat{X}_j(t)=\hat{b}_je^{-i\omega_j t}+\hat{b}^\dag e^{i\omega_j t}
\end{equation}
gives for the partial trace over Eq. (5) of the main paper 
\begin{align}\label{effeq1}
\frac{d\rho_\mathrm{m}}{dt}=&-i\sum_{j,k=1}^2g_j\alpha_k\left[e^{-i\Delta_k t}\left(\hat{X}_j(t){}_\mathrm{o}\langle1|\rho|0\rangle_\mathrm{o}-{}_\mathrm{o}\langle 1|\rho|0\rangle_\mathrm{o}\hat{X}_j(t)\right)
+e^{i\Delta_k t}\left(\hat{X}_j(t){}_\mathrm{o}\langle 0|\rho|1\rangle_\mathrm{o}-{}_\mathrm{o}\langle 0|\rho|1\rangle_\mathrm{o}\hat{X}_j(t)\right)\right].
\end{align}
The time evolution of the appearing matrix element is governed by
\begin{align}
\frac{d{}_\mathrm{o}\langle 1|\rho|0\rangle_\mathrm{o}}{dt}
=-i\sum_{j,k=1}^2g_j\alpha_ke^{i\Delta_kt}\hat{X}_j(t)\rho_\mathrm{m}-\frac{\kappa}{2}{}_\mathrm{o}\langle 1|\rho|0\rangle_\mathrm{o}
\end{align}
which has the formal solution
\begin{align}
{}_\mathrm{o}\langle 1|\rho|0\rangle_\mathrm{o}=&-i\sum_{j,k=1}^2g_j\alpha_ke^{-\frac{\kappa}{2} t}\int_0^tdt'e^{(\kappa/2+i\Delta_k)t'}
\hat{X}_j(t)
\rho_\mathrm{m}.
\end{align}
Assuming $\rho_\mathrm{m}$ to be slowly varying, we find
\begin{align}\label{matelementsolution}
{}_\mathrm{o}\langle 1|\rho|0\rangle_\mathrm{o}=&-i\sum_{j,k=1}^2g_j\alpha_k\left(\frac{\hat{b}_j^\dag e^{i(\Delta_k+\omega_j)t}}{\kappa/2+i(\Delta_k+\omega_j)}+\frac{\hat{b}_je^{i(\Delta_k-\omega_j)t}}{\kappa/2+i(\Delta_k-\omega_j)}\right)
\rho_\mathrm{m}.
\end{align}
Plugging expression (\ref{matelementsolution}) into (\ref{effeq1}) yields after some algebra and dropping off-resonant terms
\begin{align}
\frac{d\rho_\mathrm{m}}{dt}=&\sum_{j,k=1}^2G_j^2\left[\frac{\hat{b}_j^\dag\rho_\mathrm{m}\hat{b}_j-\rho_\mathrm{m}\hat{b}_j\hat{b}_j^\dag}{\kappa/2-i(\Delta_k+\omega_j)}+\frac{\hat{b}_j\rho_\mathrm{m}\hat{b}_j^\dag-\rho_\mathrm{m}\hat{b}_j^\dag\hat{b}_j}{\kappa/2-i(\Delta_k-\omega_j)}-\frac{\hat{b}_j\hat{b}_j^\dag\rho_\mathrm{m}-\hat{b}_j^\dag\rho_\mathrm{m}\hat{b}_j}{\kappa/2+i(\Delta_k+\omega_j)}-\frac{\hat{b}_j^\dag\hat{b}_j\rho_\mathrm{m}-\hat{b}_j\rho_\mathrm{m}\hat{b}_j^\dag}{\kappa/2+i(\Delta_k-\omega_j)}\right]
\\
&+G_1G_2\left[-\frac{\hat{b}_1^\dag\hat{b}_2\rho_\mathrm{m}-\hat{b}_2\rho_\mathrm{m}\hat{b}_1^\dag}{\kappa/2+i(\Delta_2-\omega_2)}+\frac{\hat{b}_1\rho_\mathrm{m}\hat{b}_2^\dag-\rho_\mathrm{m}\hat{b}_2^\dag\hat{b}_1}{\kappa/2-i(\Delta_2-\omega_2)}
-\frac{\hat{b}_2\hat{b}_1^\dag\rho_\mathrm{m}-\hat{b}_1^\dag\rho_\mathrm{m}\hat{b}_2}{\kappa/2+i(\Delta_2+\omega_1)}+\frac{\hat{b}_2^\dag\rho_\mathrm{m}\hat{b}_1-\rho_\mathrm{m}\hat{b}_1\hat{b}_2^\dag}{\kappa/2-i(\Delta_2+\omega_1)}\right.\\
&\qquad\quad\left.-\frac{\hat{b}_1\hat{b}_2^\dag\rho_\mathrm{m}-\hat{b}_2^\dag\rho_\mathrm{m}\hat{b}_1}{\kappa/2+i(\Delta_1+\omega_2)}+\frac{\hat{b}_1^\dag\rho_\mathrm{m}\hat{b}_2-\rho_\mathrm{m}\hat{b}_2\hat{b}_1^\dag}{\kappa/2-i(\Delta_1+\omega_2)}
-\frac{\hat{b}_2^\dag\hat{b}_1\rho_\mathrm{m}-\hat{b}_1\rho_\mathrm{m}\hat{b}_2^\dag}{\kappa/2+i(\Delta_1-\omega_1)}+\frac{\hat{b}_2\rho_\mathrm{m}\hat{b}_1^\dag-\rho_\mathrm{m}\hat{b}_1^\dag\hat{b}_2}{\kappa/2-i(\Delta_1-\omega_1)}\right].
\end{align}
Re-ordering terms and separating the unitary evolution from dissipative dynamics gives Eq. (12) from the main paper.